\documentclass[aps,twocolumn,showpacs,amsmath,amssymb,prc]{revtex4-1}
\usepackage{graphicx}

\begin{document}


\title{Isomeric states close to doubly magic $^{132}$Sn studied with JYFLTRAP}


\author{A. Kankainen}
\email{anu.k.kankainen@jyu.fi}
\affiliation{Department of Physics, P.O. Box 35 (YFL), FI-40014 University of Jyv\"askyl\"a, Finland}
\author{J. Hakala}
\affiliation{Department of Physics, P.O. Box 35 (YFL), FI-40014 University of Jyv\"askyl\"a, Finland}
\author{T. Eronen}
\altaffiliation{Max-Planck-Institut f\"ur Kernphysik, Saupfercheckweg 1, D-69117 Heidelberg, Germany}
\affiliation{Department of Physics, P.O. Box 35 (YFL), FI-40014 University of Jyv\"askyl\"a, Finland}
\author{D. Gorelov} 
\affiliation{Department of Physics, P.O. Box 35 (YFL), FI-40014 University of Jyv\"askyl\"a, Finland}
\author{A. Jokinen} 
\affiliation{Department of Physics, P.O. Box 35 (YFL), FI-40014 University of Jyv\"askyl\"a, Finland}
\author{V.S. Kolhinen} 
\affiliation{Department of Physics, P.O. Box 35 (YFL), FI-40014 University of Jyv\"askyl\"a, Finland}
\author{I.D. Moore} 
\affiliation{Department of Physics, P.O. Box 35 (YFL), FI-40014 University of Jyv\"askyl\"a, Finland}
\author{H. Penttil\"a} 
\affiliation{Department of Physics, P.O. Box 35 (YFL), FI-40014 University of Jyv\"askyl\"a, Finland}
\author{S. Rinta-Antila} 
\affiliation{Department of Physics, P.O. Box 35 (YFL), FI-40014 University of Jyv\"askyl\"a, Finland}
\author{J. Rissanen} 
\affiliation{Department of Physics, P.O. Box 35 (YFL), FI-40014 University of Jyv\"askyl\"a, Finland}
\altaffiliation{Nuclear Science Division, Lawrence Berkeley National Laboratory, Berkeley, California 94720, USA}
\author{A. Saastamoinen} 
\altaffiliation{Cyclotron Institute, Texas A\&M University, College Station, TX, 77843-3366, USA}
\affiliation{Department of Physics, P.O. Box 35 (YFL), FI-40014 University of Jyv\"askyl\"a, Finland}
\author{V. Sonnenschein}
\affiliation{Department of Physics, P.O. Box 35 (YFL), FI-40014 University of Jyv\"askyl\"a, Finland}
\author{J. \"Ayst\"o}
\altaffiliation{Helsinki Institute of Physics, P.O. Box 64, FI-00014 University of Helsinki, Finland}
\affiliation{Department of Physics, P.O. Box 35 (YFL), FI-40014 University of Jyv\"askyl\"a, Finland}


\date{\today}

\begin{abstract}
The double Penning trap mass spectrometer JYFLTRAP has been employed to measure masses and excitation energies for $11/2^-$ isomers in $^{121}$Cd, $^{123}$Cd, $^{125}$Cd and $^{133}$Te, for $1/2^-$ isomers in $^{129}$In and $^{131}$In, and for $7^-$ isomers in $^{130}$Sn and $^{134}$Sb. These first direct mass measurements of the Cd and In isomers reveal deviations to the excitation energies based on results from beta-decay experiments and yield new information on neutron- and proton-hole states close to $^{132}$Sn. A new excitation energy of $144(4)$~keV has been determined for $^{123}$Cd$^m$. A good agreement with the precisely known excitation energies of $^{121}$Cd$^m$, $^{130}$Sn$^m$, and $^{134}$Sb$^m$ has been found.
\end{abstract}


\maketitle

\section{Introduction}
\label{sec:intro}
Nuclei near the doubly magic $^{132}$Sn offer an interesting playground for the study of single-particle wavefunctions and their eigenvalues. Due to the $\nu 1h_{11/2}$ and $\pi 1g_{9/2}$ shells lying close to low-$j$ shells, nuclei close to $^{132}$Sn have typically long-living isomeric states (see Table~\ref{tab:prop}). Information on these isomers has been scarce since many of them decay via beta decay and the excitation energies have been based on differences in beta-decay energies. In the work presented in this paper, we have applied state-of-the-art cleaning methods at the JYFLTRAP double Penning trap mass spectrometer to study the isomers and corresponding single-particle states around $^{132}$Sn.

The isomers studied in this work have spins and parities of $11/2^-$, $1/2^-$ and $7^-$ (see Table~\ref{tab:prop}). Even-Z, odd-N nuclei just below the closed neutron shell $N=82$ favor pairing in the $\nu 1h_{11/2}$ shell. They typically have $3/2^+$ or $1/2^+$ ground states and $11/2^-$ isomeric states corresponding to $\nu 2d_{3/2}$, $\nu 3s_{1/2}$ and $\nu 1h_{11/2}$ neutron-hole states, respectively. All odd-In ($Z=49$) isotopes have $9/2^+$ ground states and $1/2^-$ isomeric states due to $(\pi 1g_{9/2})^{-1}$ and $(\pi 2p_{1/2})^{-1}$ configurations. The evolution of the excitation energies of the $11/2^-$ and $1/2^-$ isomeric states towards the closed neutron shell at $N=82$ gives us information on the neutron- and proton-hole states close to $^{132}$Sn. Even-even and odd-odd nuclei also have isomeric states near $^{132}$Sn. In this work, we have focused on the even-even nucleus $^{130}$Sn and odd-odd nucleus $^{134}$Sb both having a $7^-$ isomer. These nuclei also offer a possibility to check the consistency of our measurements since they are already well-known via $\gamma$-spectroscopy. Cd and In isotopes have also high-lying, high-spin ($I\geq17/2$) isomers which usually have relatively short half-lives (see e.g. Refs.~\cite{Sim11,Sch04,Hel03,Gen03}). Short-lived isomers are not of interest for this work focusing on isomers with half-lives longer than 100~ms.

The astrophysical $r$-process (see e.g. Ref.~\cite{Arn07}) proceeds through the region around $^{132}$Sn. The beta-decay properties and masses of the involved nuclei should be precisely known in order to more reliably compare the calculated $r$-process abundances to the observations. Isomers must also be taken into account in the $r$-process modeling. Namely, isomers and low-lying excited states can be thermally populated if the $r$-process operates at high temperatures, and thus, the beta-decay rates can significantly differ from the terrestrial rates \cite{Arn07}. Furthermore, if the $r$-process operates at such low temperatures that thermal equilibrium cannot be achieved, it becomes necessary to independently describe the population of different isomers after neutron capture and the rates for neutron capture and decay of each isomer. Precision mass measurements of both ground and isomeric states yield more accurate mass values for $r$-process calculations since the measured states have been identified. They are also important to help constrain the residual interaction used in shell-model calculations.

\begin{table}
\caption{\label{tab:prop} Properties of the nuclides studied in this work. The values are based on Ref.~\cite{nubase} unless stated otherwise. Given are half-lives $T_{1/2}$, spin-parities $I^\pi$, and excitation energies of the isomeric states $E_x$. The parenthesis in the third column indicates uncertain values of spin and/or parity. The values estimated from systematic trends from neighboring nuclides with the same Z and N parities are denoted by '\#'.}
\begin{ruledtabular}
\begin{tabular}{llll}
Nuclide & $T_{1/2}$ & $I^\pi$ & $E_x$ (keV)\\
\hline
$^{121}$Cd & $13.5(3)$~s& $(3/2^+)$ & \\
$^{121}$Cd$^m$ & $8.3(8)$~s & $(11/2^-)$ & $214.86(15)$ \\
$^{123}$Cd & $2.10(2)$~s &$(3/2)^+$ & \\
$^{123}$Cd$^m$ & $1.82(3)$~s & $(11/2^-)$ & $316.52(23)$ \\
$^{125}$Cd & $650(20)$~ms &$3/2^+$\# & \\
$^{125}$Cd$^m$ & $570(90)$~ms & $11/2^-$\# & $50(70)$ \\
$^{129}$In & $611(4)$~ms & $9/2^+$\# & \\
$^{129}$In$^m$ & $1.23(3)$~s & $1/2^-$\# & $370(40)$\footnotemark[1]\\
$^{129}$In$^n$ & $670(100)$~ms\footnotemark[2] & $23/2^-$ & $1630(56)$\footnotemark[2] \\
$^{129}$In$^p$ & $8.5(8)$~$\mu$s & $17/2^-$ & $1688.0(5)$ \\
$^{131}$In & $280(30)$~ms & $(9/2^+)$ & \\
$^{131}$In$^m$ & $350(50)$~ms & $(1/2^-)$ & $302(32)$\footnotemark[3] \\
$^{131}$In$^n$ & $320(60)$~ms & $(21/2^+)$\# & $3764(88)$\footnotemark[3] \\
$^{130}$Sn & $3.72(7)$~min & $0^+$ &  \\
$^{130}$Sn$^m$ & $1.7(1)$~min & $7^-$\# & $1946.88(10)$ \\
$^{131}$Sn & $56.0(5)$~s & $(3/2^+)$ &  \\
$^{131}$Sn$^m$ & $58.4(5)$~s & $(11/2^-)$ & $69(14)$\footnotemark[3] \\
$^{132}$Sb & $2.79(5)$~min & $(4^+)$ & \\
$^{132}$Sb$^m$ & $4.15(5)$~min & $(8^-)$ & 200(30)\\
$^{134}$Sb & $780(60)$~ms & $(0^-)$ & \\
$^{134}$Sb$^m$ & $10.07(5)$~s\footnotemark[4] & $(7^-)$ & $279(1)$\footnotemark[4] \\
$^{133}$Te & $12.5(3)$~min & $(3/2^+)$ &\\
$^{133}$Te$^m$& $55.4(4)$~min & $(11/2^-)$ & $334.26(4)$ \\
\end{tabular}
\footnotetext[1]{The value is a weighted mean of the results from \cite{Ale78,Spa87,Gau04}.}
\footnotetext[2]{The value has been taken from \cite{Gau04}.}
\footnotetext[3]{The value has been taken from \cite{Fog04}.}
\footnotetext[4]{The value has been taken from \cite{She05}.}
\end{ruledtabular}
\end{table}

\section{Experimental method}
\label{sec:exp}

The ions of interest were produced via fission reactions induced by 25-MeV protons on $\rm ^{nat}U$ or $^{232}$Th at the Ion Guide Isotope Separator On-Line (IGISOL) facility \cite{Ays01}. A 15-mg/cm$^2$-thick uranium target foil was used to produce neutron-rich $^{121-128}$Cd, $^{131}$In, $^{130-134}$Sn, and $^{132-140}$Te in June 2009. $^{129}$In and $^{131-136}$Sb isotopes were produced with a 14-mg/cm$^2$-thick $^{232}$Th target in a second experiment in May 2010. The results concerning the measured ground-state mass values have already been published in Ref.~\cite{Hak12}. This paper focuses on the isomeric states.

The fission products are stopped in the ion-guide gas cell filled with helium at a pressure of around 200~mbar. There, a good fraction of ions ends up as singly-charged via charge-exhange reactions. The ions are extracted from the gas cell by differential pumping and with a sextupole ion guide (SPIG) \cite{Kar08}. After acceleration to 30 kV and mass-separation with a 55$^\circ$ dipole magnet, the continuous ion beam with a selected mass number $A$ will be sent to a gas-filled radio-frequency quadrupole cooler and buncher (RFQ) \cite{Nie01}. The RFQ cools the ions and injects them as narrow ion bunches to the JYFLTRAP double Penning trap \cite{Kol04,Ero12}. 

JYFLTRAP consists of two cylindrical Penning traps inside a $7$-T superconducting solenoid. The first trap, the purification trap, is used for beam purification via a mass-selective buffer-gas cooling technique \cite{Sav91}. With a typical mass resolving power of around $m/\Delta m \approx 3\times 10^4$ neighbouring isobars can usually be separated before extracting them through a narrow diaphragm towards the second trap, the precision trap. There, high-precision mass measurements are performed employing the time-of-flight ion cyclotron resonance (TOF-ICR) technique \cite{Gra80,Kon95}. 

The ions in a Penning trap have three different eigenmotions: axial, magnetron and reduced cyclotron motions with frequencies $\nu_z$, $\nu_-$ and $\nu_+$, respectively. According to the invariance theorem \cite{Bro86}, the sideband frequency $\nu_-+\nu_+$ corresponds to the true cyclotron frequency with a high precision even in a non-ideal Penning trap: 

\begin{equation}
\label{eq:1}
\nu_c = \frac{1}{2\pi}\frac{q}{m}B,
\end{equation}
where $B$ is the magnetic field, and $q$ and $m$ are the charge and the mass of the ion. In the precision trap, a dipole excitation at magnetron motion frequency is firstly applied to increase the magnetron radius for all ions. A subsequent quadrupole excitation is used to convert the magnetron motion into a reduced cyclotron motion periodically. When the excitation frequency matches the sideband frequency $\nu_-+\nu_+$, the radial energy of the ions reaches its maximum as the initially pure magnetron motion is fully converted to reduced cyclotron motion. The gain in radial energy is observed as a shorter time-of-flight to the microchannel plate detector (MCP) when the ions are extracted from the trap in the strong magnetic field gradient.

The magnetic field $B$ is calibrated with a reference whose atomic mass $m_{ref}$ is well-known. In this experiment, $^{130}$Xe ($m=129.903509351(15)$~u \cite{Red09a}) was used as a reference except for $^{131}$Sn, for which $^{132}$Xe ($m=131.904155086(10)$~u \cite{Red09b}) was employed. Since singly-charged ions were used, the mass of the nuclide of interest can be determined as:

\begin{equation}
\label{eq:2}
m_{meas}=r(m_{ref}-m_e)+m_e,
\end{equation}
where $r=\frac{\nu_{c,ref}}{\nu_{c,meas}}$ is the measured cyclotron frequency ratio between the reference ion and the ion of interest, and $m_e$ is the electron mass.

Since the studied isomeric states lie close to the ground states, they could not be fully resolved by employing only the purification trap. A so-called Ramsey cleaning technique \cite{Ero08} was applied for resolving the isomers. There, the purified ions from the first trap are further cleaned by applying a dipolar excitation at reduced cyclotron frequency of the contaminant ion in the form of time-separated oscillatory fields \cite{Ram90,Bol92,Kre07} in the precision trap. In this way, the unwanted species are driven to a larger cyclotron orbit but the ions of interest are unaffected. After the dipolar excitation, only the ions of interest can pass through the 2-mm diaphragm back to the purification trap for recooling and recentering before the actual mass measurement in the precision trap. With this additional cleaning method, a mass resolving power up to $m/\Delta m\approx 10^6$ can be achieved. 

\begin{figure}
\centering
\begin{tabular}{c}
\includegraphics[width=0.45\textwidth,clip]{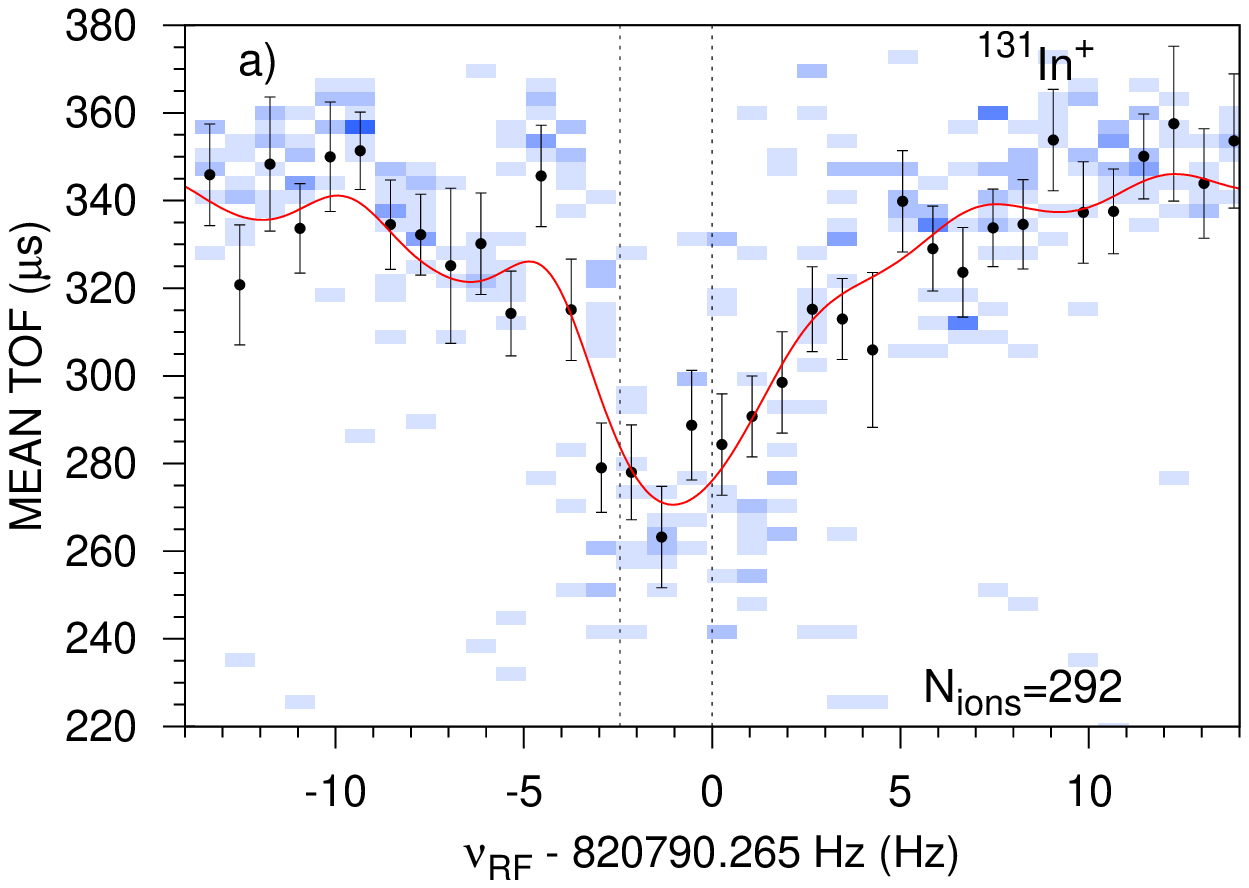}\\
\includegraphics[width=0.45\textwidth,clip]{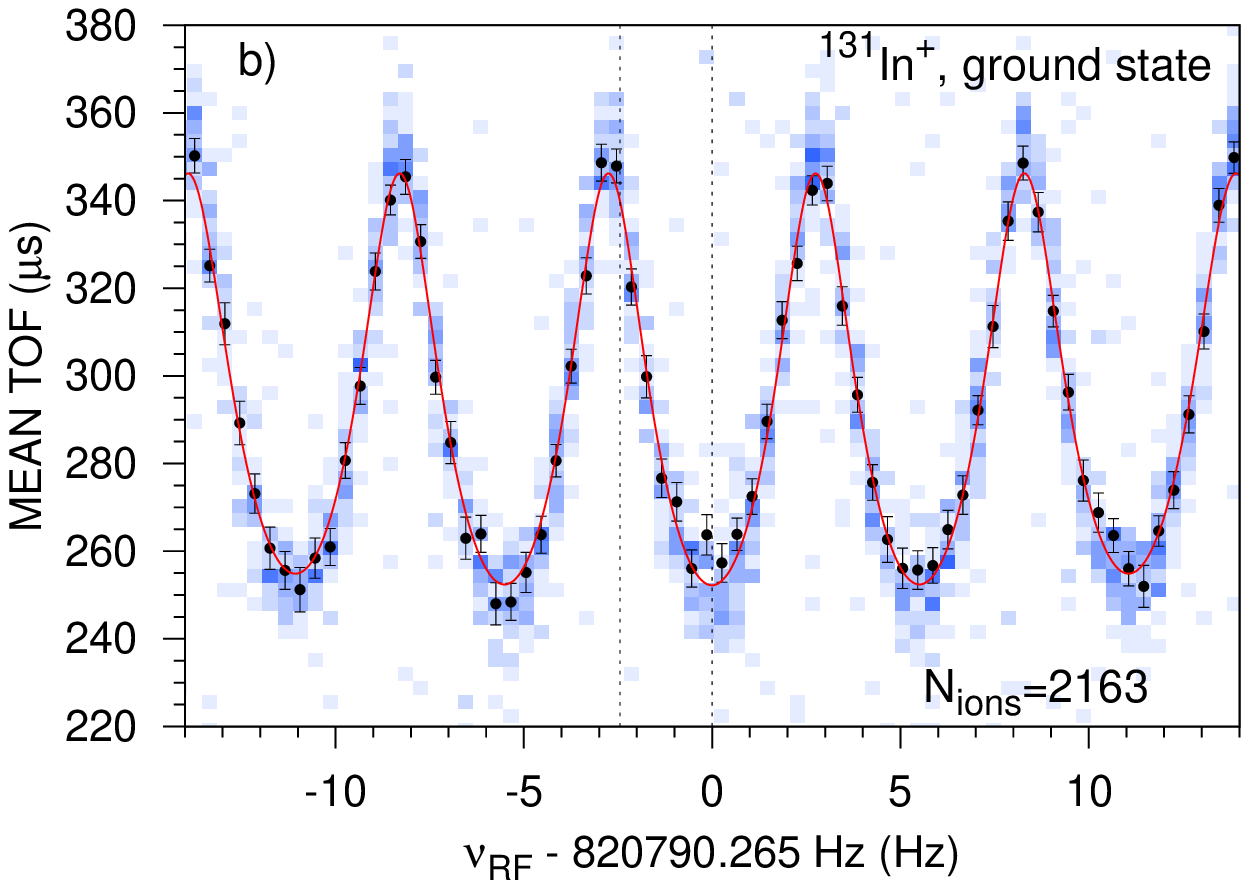}\\
\includegraphics[width=0.45\textwidth,clip]{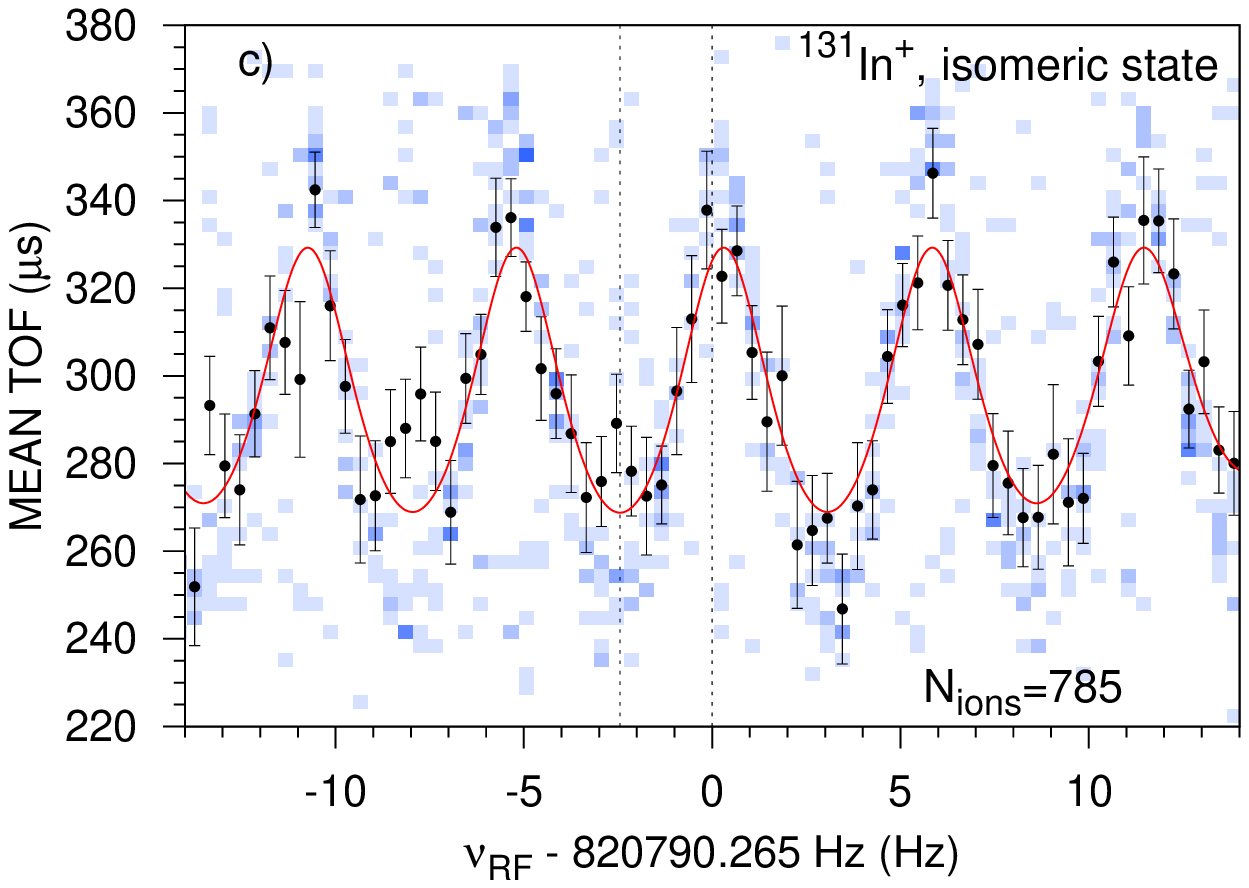}
\end{tabular}
\caption{(Color online) TOF-ICR spectra of $^{131}$In: a) Without Ramsey cleaning and with a continuous quadrupolar RF excitation of 200~ms, b) Ramsey-cleaned ground state with a 25-150-15~ms (On-Off-On) excitation and c) Ramsey-cleaned isomeric state with a 25-150-15~ms (On-Off-On) excitation. The number of ions in a bunch has been limited to $1-2$~ions/bunch and the time gate has been set to $217.6-384.0~\mu s$ in these figures. The blue squares indicate the number of ions in each time-of-flight bin: the darker the color, the more ions there are. The dashed lines show the positions of the resonance frequencies for the ground and isomeric states.}
\label{fig:TOF}       
\end{figure}

In this work, a dipolar excitation pattern of $20-40-20$~ms (On-Off-On) was used for Cd isotopes, $15-30-15$~ms (On-Off-On) for $^{130}$Sn, $20-50-20$~ms (On-Off-On) for $^{131}$Sn and $10-40-10$~ms (On-Off-On) for the rest. Time-separated oscillatory fields were also applied for the quadrupole excitation in the precision trap and the resulting frequency spectrum was fitted with the theoretical lineshape \cite{Kre07,Geo07}. An excitation pattern of $25-350-25$~ms (On-Off-On) was used except for In and Sn isotopes. The time between the 25-ms pulses was shortened to 150~ms for In isotopes and increased to 725~ms for Sn isotopes. Figure~\ref{fig:TOF} shows examples of TOF-ICR spectra with and without Ramsey cleaning and excitation for $^{131}$In.

The data were collected interleavedly \cite{Ero09}: after one or two frequency sweeps for the reference ion, a few frequency sweeps were collected for the ion of interest and this pattern was repeated as long as required for sufficient statistics (typically for few hours). The interleaved scanning reduces the uncertainty due to time-dependent fluctuations in the magnetic field, which for JYFLTRAP has been determined to be $\delta_B(\nu_{ref})/\nu_{ref}=5.7(8)\times 10^{-11}\text{min}^{-1}\Delta t$, where $\Delta t$ is the time between the two reference measurements. The data files were split into smaller parts in such a way that a proper count-rate class analysis \cite{Kel03} was possible at least for the reference ion. If the count-rate class analysis was not possible, the number of ions was limited to $1-2$~ions/bunch and the uncertainty of the obtained cyclotron frequency was multiplied by a coefficient deduced from the closest neighbouring isotope possible. Frequency ratios were calculated for each data pair. The error due to the time-dependent magnetic field fluctuation was quadratically added to the statistical uncertainty of each frequency ratio. The weighted mean of the measured frequency ratios was calculated and used as the final value. The inner and outer errors \cite{Bir32} of the data sets were compared and the larger value of these two was taken as the error of the mean. Finally, the uncertainty due to the mass-dependent shift $\delta_{m,lim}(r)/r=(7.5 \pm 0.4 \times 10^{-10}/u)\times \Delta m$ \cite{Elo09} and an additional residual relative uncertainty $\delta_{res,lim}(r)/r=7.9\times 10^{-9}$ \cite{Elo09} were quadratically added to the error. The obtained frequency ratios are collected in Table~\ref{tab:ratios}.

\section{Results and discussion}
\label{sec:results}

\begin{table*}
\caption{\label{tab:ratios} Cyclotron frequency ratios ($r=\nu_{c,ref}/\nu_{c,meas}$), resulting mass-excess values (in keV) and comparison to the literature values \cite{nubase} for the measured isomers. $^{130}$Xe was used as a reference except for $^{131}$Sn for which $^{132}$Xe was employed.}
\begin{ruledtabular}
\begin{tabular}{llllll}
Isomer & $r$ & $\rm ME_{JYFL}$ (keV) & $\rm E_{x,JYFL}$ (keV) & $ \rm E_{x,LIT}$ (keV) & $\rm \Delta E_{x,JYFL-LIT}$ (keV) \\
\hline
$^{121}$Cd$^m$ & $0.930792072(22)$ & $-80858.7(26)$ & $215(4)$ & $214.86(15)$ & $1(4)$ \\
$^{123}$Cd$^m$ & $0.946217831(25)$ & $-77271(3)$ & $144(4)$ & $316.52(23)$ & $-173(4)$\\
$^{125}$Cd$^m$ & $0.961647897(26)$ & $-73162(4)$ & $186(5)$ & $50(70)$ & $140(70)$\\
$^{129}$In$^m$ & $0.992446580(27)$ & $-72379(4)$ & $459(5)$ & $370(40)$\footnotemark[1] & $90(40)$ \\
$^{131}$In$^m$ & $1.007881690(59)$ & $-67660(8)$ & $365(8)$ & $302(32)$\footnotemark[2] & $63(33)$\\
$^{130}$Sn$^m$ & $1.000096653(24)$ & $-78185(3)$ & $1948(5)$ & $1946.88(10)$ & $1(5)$\\
$^{131}$Sn$^x$ & $0.992516774(26)$ & $-77230(4)$\footnotemark[3] & - & $69(14)$\footnotemark[2] &\\
$^{134}$Sb$^m$ & $1.030925604(24)$ & $-73740.0(29)$ & $281(4)$ & $279(1)$\footnotemark[4] & $2(4)$ \\
$^{133}$Te$^m$ & $1.023154364(20)$ & $-82595.8(24)$ & $342(4)$ & $334.26(4)$ & $8(4)$\\
\end{tabular}
\footnotetext[1]{The value is a weighted mean of the results from \cite{Ale78,Spa87,Gau04}.}
\footnotetext[2]{The value has been taken from \cite{Fog04}.}
\footnotetext[3]{The value is the measured value. The value corrected due to an unknown mixture of states is $-77262(20)$~keV when $\rm E_x=65.1$~keV \cite{Fog04} is assumed for the isomer.}
\footnotetext[4]{The value has been taken from \cite{She05}.}
\end{ruledtabular}
\end{table*}

\subsection{$11/2^-$ isomers}
$11/2^-$ isomers are typical for even-$Z$, odd-$N$ nuclei below the $N=82$ neutron shell. For example, long-lived $11/2^-$ isomers are found in $^{119,121,123,125}$Cd ($Z=48$), $^{123,125,127,129,131,133}$Te ($Z=52$) and $^{129,131,133,135}$Xe ($Z=54$). Typical excitation energies are around $100-400$~keV (see Fig.~\ref{fig:Cd_syst}) and a sharp rise is observed when approaching the closed $N=82$ shell. In odd-$N$ Sn isotopes the $3/2^+$ and $11/2^-$ states are very close to each other and the order changes from isotope to isotope. The increase in the excitation energy towards the $N=82$ shell is also observed in Sn isotopes after $N=77$.

\subsubsection{$^{121}$Cd$^m$}
\label{sec:cd121}
$^{121}$Cd has a $3/2^+$ ground state with a half-life of $13.5(3)$~s \cite{Wei65,Gra74,Sch74,Fog82a} and a $11/2^-$ isomeric state with a half-life of $8.3(8)$~s \cite{Fog82a}. The excitation energy of the isomer has been determined as $70(170)$~keV from the difference between the beta-decay energies of the ground and isomeric states \cite{Ale82}. No direct evidence for the excitation energy was found from the $\gamma$-$\gamma$ coincidence data \cite{Fog82b}. However, by assuming that there should be common levels de-exciting both to the ground and to the isomeric state, three common levels could be found and an excitation energy of $214.89$~keV was deduced for the isomer \cite{Fog82b}. The JYFLTRAP value, $215(4)$~keV, is the first direct measurement of the excitation energy of this isomer and it confirms the deductions made in Ref.~\cite{Fog82b}.

\subsubsection{$^{123}$Cd$^m$}
\label{sec:cd123}
The existence of two long-lived states was not observed in the first measurements on $^{123}$Cd \cite{Ree83,Gok86,Hof86} since the half-lives are so similar. The half-life for the $11/2^-$ isomer, $1.82(3)$~s, was determined in Refs.~\cite{Mac86,Huc89a}. The excitation energy of the isomer, $316.52(23)$~keV \cite{Huc89b}, is not based on observations of $\gamma - \gamma$ coincidences or on a direct measurement of $\gamma$-transition energy. It was searched for by maximizing the number of shared levels de-exciting to the ground and isomeric states. Two shared levels, at $1061$~keV and $2240$~keV, were found when the $\gamma$-energy range of $260-400$~keV was applied in the search. The JYFLTRAP result, $\rm E_x=144(4)$~keV, is below the energy range used for the search and is in a clear disagreement with Ref.~\cite{Huc89b}. Based on our result, the most probable shared level would be the state at $263.87(2)$~keV \cite{Huc89b} de-excited by $263.87(2)$~keV and $123.67(6)$~keV $\gamma$-transitions to the ground and isomeric states, respectively. This yields an energy of $140.20(6)$~keV for the $11/2^-$ state. Since the $11/2^-$ state is below $263.87(2)$~keV, a spin assignment of $7/2^+$ is possible for the state at $263.87(2)$~keV. The spin assignment ($7/2^+$) is further supported by the beta-decay $\log ft$ value of $5.09$ from the ($7/2^+$) state in $^{123}$Ag as well as by the half-life of $80(15)$~ns, which is compatible with an $E2$ transition to the $3/2^+$ ground state with a hindrance factor of 4 \cite{Huc89b}. On the other hand, the intensity ratio of the $124$-keV and $264$-keV $\gamma$-rays ($16$~\% \cite{Huc89b}) should be much smaller if $M2$ and $E2$ transitions are assumed. 

\begin{figure}
\center{
\resizebox{0.45\textwidth}{!}{
\includegraphics{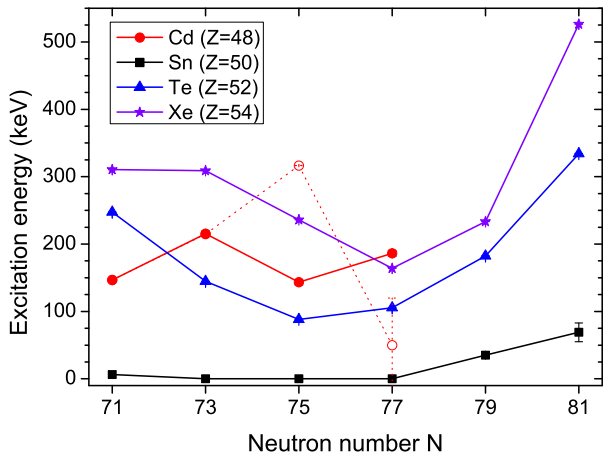}}}
\caption{(Color online) Systematics of the $11/2^-$ states in even-$Z$, odd-$N$ nuclei. The open red circles and the dotted red line show the trend for the Cd isotopes with previously adopted excitation energies \cite{Spa87,Huc89b}.}
\label{fig:Cd_syst} 
\end{figure}

The energy systematics of the $11/2^-$ states in even-$Z$, odd-$N$ isotopes below the $N=82$ neutron shell supports an excitation energy of around $140$~keV for the isomer and the adopted value at $317$~keV is clearly off the trend (see Fig.~\ref{fig:Cd_syst}). The new JYFLTRAP result also shifts the energy levels above the $11/2^-$ isomer towards lower energies. The energy levels of the $15/2^-$, $19/2^-$, $21/2^-$, and $23/2^-$ states in the band built on the $11/2^-$ isomer in $^{121}$Cd and $^{123}$Cd \cite{Hwa02} are much closer to each other when the JYFLTRAP value is applied. 

\begin{figure}
\center{
\resizebox{0.45\textwidth}{!}{
\includegraphics{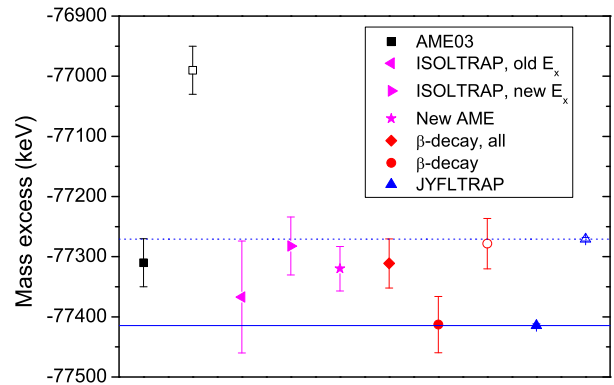}}}
\caption{(Color online) Comparison of different results for the $^{123}$Cd ground state (full symbols) and isomeric state (open symbols). The ISOLTRAP data have been corrected with the old ($E_x=316.52(23)$~keV \cite{Huc89b}) and new excitation energy ($E_x=144(4)$~keV, this work), respectively. The AME03 value \cite{AME03} as well as the new AME value given in Ref.~\cite{Bre10} are mainly based on the beta-decay result of Ref.~\cite{Spa87} assuming only one $\beta$-decaying state. The results of this work agree with the $\beta$-decay data \cite{Spa87,Huc89a} when the ground and isomeric state are treated separately. However, JYFLTRAP disagrees with the AME03 and the updated ISOLTRAP value.}
\label{fig:cd123} 
\end{figure}

$^{123}$Cd has also been studied at ISOLTRAP but the ground and isomeric states could not be resolved \cite{Bre10}. Thus, the measured mass-excess value of $-77210(25)$~keV was corrected with $\rm E_x=316.52(23)$~keV \cite{Huc89b} to obtain the ground-state mass value of $-77367(93)$~keV \cite{Bre10}. As the ISOLTRAP value is higher than the JYFLTRAP value for the isomer ($-77271(3)$~keV), it most likely belongs to the isomeric state (see Fig.~\ref{fig:cd123}). 

The mass evaluation performed in Ref.~\cite{Bre10} yielded a new adjusted value of $-77320(37)$~keV for the ground state of $^{123}$Cd. However, most of the influence in the mass evaluation still came from a $\beta$-endpoint measurement of $^{123}$Cd \cite{Spa87}. There, six different $\gamma$-transition gates based on the results of Ref.~\cite{Hof86} were used for determining the endpoint energies for the ground-state $\beta$-decay. The obtained value of $Q_\beta=6115(33)$~keV \cite{Spa87} results in a ground-state mass-excess value of $-77311(41)$~keV, around $100$~keV higher than the JYFLTRAP value. 

In a newer $\beta$-decay experiment \cite{Huc89a} it was shown that many of the gating transitions used in Ref.~\cite{Spa87} do not belong to the ground-state $\beta$-decay but to the decay from the isomeric state. The used $428$-keV, $2461$-keV and $2602$-keV $\gamma$-transitions belong to the $^{123}$Cd$^m$ $\beta$-decay whereas the $1831$-keV and $1843$-keV $\gamma$-transitions result from the ground-state $\beta$-decay. The used $1695$-keV gate is predominantly fed by the ground-state decay but there is also a small contribution from the isomeric state. The weighted mean based on the endpoint energies belonging to the $^{123}$Cd$^m$ $\beta$-decay is $Q_\beta=6148(35)$~keV. Similarly, the $Q_\beta$ value for the ground-state $\beta$-decay is $6013(41)$~keV. Thus, an excitation energy of $135(53)$~keV is obtained for the $11/2^-$ isomer in $^{123}$Cd based on the beta-decay studies \cite{Spa87,Huc89a}. This strongly supports the new JYFLTRAP value, also seen in Fig.~\ref{fig:cd123}.

\subsubsection{$^{125}$Cd$^m$}
\label{sec:cd125}

The first $\beta$-decay experiments on $^{125}$Cd did not resolve the ground and isomeric states \cite{Gok86,Hof86,Spa87}. The half-lives for both states were determined in Refs.~\cite{Mac86,Huc89a}. Similarly to $^{123}$Cd, three out of the four $\gamma$-transitions ($191$-~, $262$-, and $1614$-keV transitions) used to gate the beta spectra in Ref.~\cite{Spa87} turned out to belong to the isomeric $\beta$-decay and not to the assumed ground-state decay \cite{Huc89a}. The difference in the beta-decay energies gives an estimate of $50(80)$~keV for the isomer. The JYFLTRAP result, $188(5)$~keV, is the first direct measurement of the excitation energy of the $11/2^-$ isomer. With the new JYFLTRAP values, the trend between the $N=73-75$ odd-$N$ isotopes is very similar for Cd ($Z=48$) and Te ($Z=52$) as shown in Fig.~\ref{fig:Cd_syst}. Recently, more attention has been paid on the $\mu s$ isomers in $^{125}$Cd \cite{Hel03,Sch04,Hot07,Sim11}. $\gamma$-lines de-exciting the $19/2^+$ and $15/2^-$ levels above the $11/2^-$ isomer have been observed in all recent experiments \cite{Hel03,Sch04,Hot07,Sim11}. Weak delayed $\gamma$-transitions with energies of $486$ and $667$~keV were observed only in Ref.~\cite{Hel03}. Interestingly, the difference between these transitions (181~keV) is close to the observed isomeric energy.

\subsubsection{$^{131}$Sn$^x$}
\label{sec:sn131}

For $^{131}$Sn, the isomeric state could not be resolved with JYFLTRAP. The excitation energy of the isomer was determined as $160(100)$~keV and suggested as a level at $241.8$~keV in an early $\beta$-decay study \cite{Fog84a}. A recent $\beta$-decay experiment yielded an excitation energy of $69(14)$~keV \cite{Fog04}. A more precise value of $65.1$~keV not confirmed by coincidence measurements may also be deduced from the level scheme of $^{131}$Sn \cite{Fog04}. The latter excitation energies were too low to be resolved with the current JYFLTRAP facility. $^{131}$Sn has also been studied at ISOLTRAP \cite{Sik05,Dwo08}. The measured mass-excess values ($-77242(15)$~keV \cite{Sik05} and $-77222(4)$~keV \cite{Dwo08}) agree with the value determined at JYFLTRAP ($-77230(4)$~keV). Therefore, it is likely that the same state or a similar mixture of states has been measured with both traps.  

The $\beta$-decay of the $11/2^-$ isomer in $^{131}$Sn feeds the $15/2^-$ state at $1606.7$~keV in $^{131}$Sb \cite{Huc81}. On the other hand, the $3/2^+$ ground state feeds the ($5/2^+$) state at $798.494$~keV in $^{131}$Sb \cite{Huc81}. In the future, the different $\gamma$-rays related to the isomeric ($447.4$-, $450.03$-, $1226.03$-, and $1229.23$-keV $\gamma$-rays) and ground-state $\beta$-decay ($798.494$-keV $\gamma$-rays) of $^{131}$Sn could be used to identify the isomer to ground-state ratio of the measured state. Another possibility would be the identification via laser spectroscopy or selective laser ionization. Laser spectroscopy on the ground and isomeric states of $^{131}$Sn has already been performed and laser ionization has been used to ionize the Sn isotopes \cite{LeBlanc05}. In addition, the resolving power of JYFLTRAP could be enhanced via doubly-charged or more highly-charged ions in the future. 

\subsubsection{$^{133}$Te$^m$}
\label{sec:te133}
The $11/2^-$ isomer in $^{133}$Te decays mainly via $\beta$-decay to $^{133}$I but a good fraction ($b=16.5(20)$~\%) proceeds via an $M4$ internal transition to the ground state. The isomeric transition has been studied carefully via $\gamma$-spectroscopy \cite{Alv57,Ber68,Fuj74,Bor79} and an excitation energy of $\rm E_x=334.27(4)$~keV has been determined. The JYFLTRAP result, $E_x=342(4)$~keV, deviates from the measured excitation energy by $8(4)$~keV. No clear explanation for the discrepancy has been found. A new mass measurement of $^{133}$Te$^m$ would be desirable in the future.

\subsection{$1/2^-$ isomers in $^{129}$In and $^{131}$In}
\label{sec:In}

\begin{figure}
\center{
\resizebox{0.45\textwidth}{!}{
\includegraphics{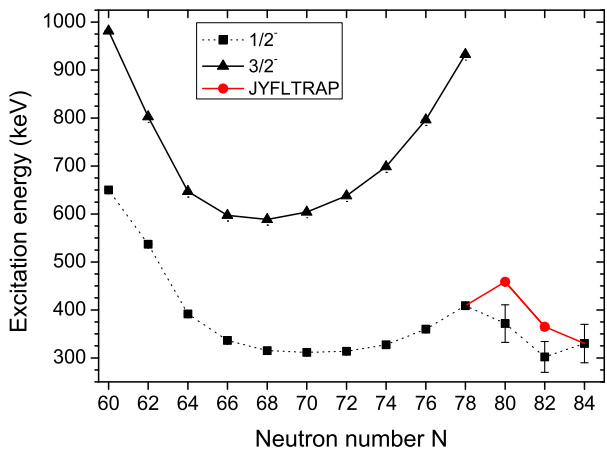}}}
\caption{(Color online) Systematics of the lowest $1/2^-$ and $3/2^-$ levels in even-$N$ In isotopes. The ground state for these isotopes is always $9/2^+$.}
\label{fig:In} 
\end{figure}

Proton-hole states close to doubly magic $^{132}$Sn can be probed by even-$N$ In ($Z=49$) isotopes. They all have $(9/2^+)$ ground states and $(1/2^-)$ isomeric states corresponding to a proton hole in the $\pi 1g_{9/2}$ or in the $\pi 2p_{1/2}$ shell. The evolution of the $1/2^-$ states in odd-$A$ In isotopes (see Fig.~\ref{fig:In}) gives us interesting information on the single-particle energies when approaching towards $N=82$.

\subsubsection{$^{129}$In}
\label{sec:in129}
Previously, the excitation energy of the $1/2^-$ isomer in $^{129}$In ($\rm E_x=200(600)$~keV \cite{Ale78}, $\rm E_x=380(70)$~keV \cite{Spa87} and $\rm E_x=369(46)$~keV \cite{Gau04}) has been based on differences in the $Q_\beta$ values between the isomeric and ground-state $\beta$-decays. In this work, JYFLTRAP yields the first directly measured value for the excitation energy of the isomer: $E_x=459(5)$~keV. It is around 90~keV higher than the values based on beta-decay experiments but taking into account the uncertainties related to the $\beta$-decay studies, the difference is not so dramatic. The new value changes the trend in the excitation energies of the $1/2^-$  isomeric states in odd-$A$ In isotopes. With the previously adopted value, the excitation energy starts to decrease already at $N=78$ but now it increases until $N=80$ (see Fig.~\ref{fig:In}). Beta-decaying ($23/2^-)$ isomer \cite{Gau04} was not searched for in the JYFLTRAP experiment, and thus could not be confirmed.

\subsubsection{$^{131}$In}
\label{sec:in131}
\begin{figure}
\center{
\resizebox{0.45\textwidth}{!}{
\includegraphics{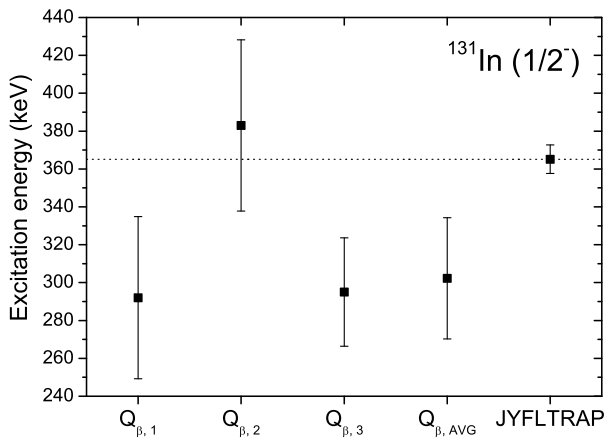}}}
\caption{The excitation energies for the $(1/2^-)$ isomer in $^{131}$In based on the difference between the $Q_\beta$ value for the $1/2^-$ isomer ($Q_\beta=9524(26)$~keV) and $Q_\beta$ values for the $9/2^+$ ground state of $^{131}$In gated by 1) 4487-keV ($Q_\beta=9232(34)$~keV), 2) 3990-keV ($Q_\beta=9141(37)$~keV) and 3) 2434-keV $\gamma$-transitions ($Q_\beta=9229(12)$~keV) \cite{Fog04}. $Q_{\beta,\rm{AVG}}$ is calculated using the weighted mean of the ground-state $\beta$-decay results \cite{Fog04}.}
\label{fig:in131_fig} 
\end{figure}

$^{131}$In has three different states with rather similar half-lives of around $300$~ms. In addition to the $(9/2^+)$ ground state and the $(1/2^-)$ isomeric state at $302(32)$~keV \cite{Fog04}, a high-spin isomer ($21/2^+$) has been observed at $3764(22)$~keV \cite{Fog04}. The JYFLTRAP measurement focused on the ground state and the $(1/2^-)$ isomer of $^{131}$In. Since the excitation energy of the high-spin isomer is almost 4~MeV, it does not interfere with the mass measurements of the other two states. The JYFLTRAP mass-excess value for the $1/2^-$ isomeric state, $E_x=365(8)$~keV, disagrees with the value based on the differences in the $Q_\beta$ values \cite{Fog04}. On the other hand, when the $Q_\beta$ results for different gating transitions \cite{Fog04} are compared with the JYFLTRAP value, an agreement is found with one of the results (see Fig.~\ref{fig:in131_fig}). 


\subsection{$7^-$ isomers in $^{130}$Sn and $^{134}$Sb}

\subsubsection{$^{130}$Sn}
\label{sec:sn130}
Even-$Z$, $N=80$ isotones, such as $^{130}$Sn, $^{132}$Te, $^{134}$Xe, $^{136}$Ba, $^{138}$Ce and $^{140}$Nd have a $7^-$ isomeric state at around $2$~MeV. This state can be explained by the $(\nu 1h_{11/2})^{-1}\otimes(\nu 2d_{3/2})^{-1}$ configuration which results in a quartet of states with spins and parities $7^-$, $6^-$, $5^-$ and $4^-$. In $^{130}$Sn, the $7^-$ state is located above the $0^+$ ground state and the $2^+$ first excited state, thus forming an yrast trap. The beta decay of the $(7^-)$ isomer was studied for the first time and an excitation energy of around 1.8~MeV was estimated in Ref.~\cite{Ker74}. Later, a more precise value of $1946.88(10)$~keV has been obtained based on $\gamma$-spectroscopy following the $\beta$-decay of $^{130}$In \cite{Fog81}. The JYFLTRAP result, $\rm E_x=1948(5)$~keV, agrees with this precise value and gives further support that the JYFLTRAP Penning trap works well for the studies of isomeric states.

The ground and isomeric states of $^{130}$Sn have also been measured at ISOLTRAP \cite{Sik05}. The ISOLTRAP value for the ground-state mass excess ($-80134(16)$~keV \cite{Sik05}) is almost in a perfect agreement with JYFLTRAP. The ISOLTRAP mass-excess value for the $7^-$ isomer ($-78190(11)$~keV \cite{Sik05}) is lower than but still in a good agreement with the JYFLTRAP value. The excitation energy of the isomer ($\rm E_x=1944(19)$~keV \cite{Sik05}) obtained at ISOLTRAP agrees well with JYFLTRAP but is less precise.

\subsubsection{$^{134}$Sb}
\label{sec:sb134}
$^{134}$Sb having one proton and neutron above the closed $Z=50$ and $N=82$ shells offers an ideal test case to study the proton-neutron interaction near doubly magic $^{132}$Sn. In this respect, $^{134}$Sb is similar to the well-studied nucleus $^{210}$Bi above the closed $Z=82$ and $N=126$ shells in the $^{208}$Pb region. Whereas the lowest energy levels in $^{210}$Bi are members of a $0^-,1^-,...,9^-$ multiplet resulting from the configuration $(\pi 1h_{9/2})^1\otimes (\nu 2g_{9/2})^1$, $^{134}$Sb has a multiplet of states $0^-,1^-,...,7^-$ corresponding to the $(\pi 1g_{7/2})^1\otimes(\nu 2f_{7/2})^1$ configuration. 

The levels of $^{134}$Sb have been intensively studied at the OSIRIS facility in Studsvik \cite{Ker72,Fog90,Kor02}. The $0^-$ ground state and the $7^-$ isomeric state were observed already in the 1970s \cite{Ker72}. In 1990, $318$-keV $\gamma$-transition was detected and assigned to a $1^-$ state in $^{134}$Sb \cite{Fog90}. Later, it was found that the $1^-$ state is located only $13$~keV above the ground state and the previously proposed $1^-$ state matches with the $2^-$ level \cite{Kor02}. In addition, $3^-$ and $4^-$ levels were identified \cite{Kor02}. The last members of the multiplet were recently observed at CERN/ISOLDE in a $\gamma$-spectroscopy experiment \cite{She05}. There, an energy of $\rm E_x=279(1)$~keV was found for the $7^-$ state based on $\gamma$-$\gamma$ coincidence relations. The JYFLTRAP value, $E_x=281(4)$~keV, agrees nicely with this precise result and confirms the position of the $7^-$ isomer. Interestingly, the shell-model calculations carried out in Ref.~\cite{Kor02} yield an excitation energy identical to the experimental value for the $7^-$ state. There, the residual interaction was based on the Kuo-Herling matrix elements deduced from the $^{208}$Pb region \cite{Cho92}.

\subsection{Isomeric to ground state intensity ratios}
\label{sec:fraction}
Before each mass measurement, TOF-ICR spectra were measured by applying continuous quadrupole excitation ($200$, $400$ or $800$~ms) in the precision trap without Ramsey cleaning. The resulting TOF-ICR spectrum is thus composed of two identical superimposed TOF-ICR curves, offset in frequency by their cyclotron frequency difference and thus their intensity ratio is obtained from the weight of the two curves. Since these spectra were only collected for setting up proper excitation schemes for the actual mass measurements, there were only a few files containing enough data to obtain the isomer to ground state intensity ratios from the resonance fits. However, a general trend was observed in the intensities: the higher-spin states dominate the resonances of Cd and In isotopes. The $(7^-)$ isomer in $^{134}$Sb was also produced in much more abundance than the $(0^-)$ ground state. 

Isomeric yield ratios for determining the angular momenta of primary fission fragments have been intensively studied at the ion-guide isotope separator on-line facility in Tohoku \cite{Tan93,Got99}. There, they have shown that the higher-spin states are typically favored in fission but on the other hand, sudden changes happen in the isomeric yield ratios, e.g. due to shell effects. This is also observed at JYFLTRAP. For instance, the intensity of the $(11/2^-)$ isomer in $^{133}$Te ($N=81$) was about the same as for the ground state whereas in $^{121}$Cd ($N=73$) it was more than four times the ground-state intensity.

\begin{figure}
\center{
\resizebox{0.45\textwidth}{!}{
\includegraphics{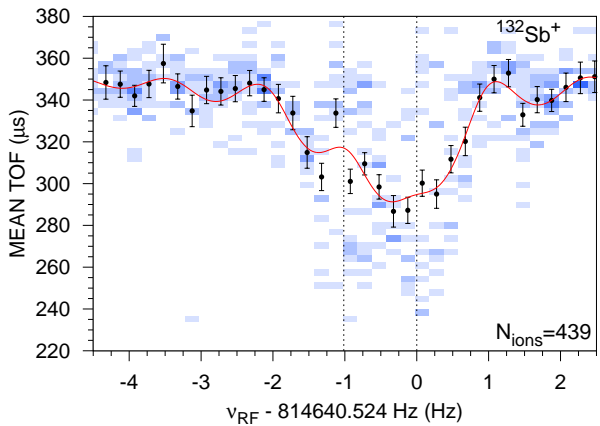}
}}
\caption{(Color online) Time-of-flight spectrum for $^{132}$Sb after a quadrupole excitation of $T_{RF}=800$~ms in the precision trap. Both the ground and the isomeric state can be seen in the spectrum. The used count window was $1-2$~ions/bunch and time gate $217.6-384.0~\mu s$.}
\label{fig:sb132}
\end{figure}

Due to lack of time, the isomeric state for the one-proton and a neutron-hole nucleus $^{132}$Sb was not measured in this work by using the Ramsey cleaning and excitation schemes as was done for the ground state \cite{Hak12}. Nevertheless, the isomeric state was observed in a collected time-of-flight spectrum with a $800$~ms quadrupole excitation (see Fig.~\ref{fig:sb132}). The frequency ratio between the isomer and the ground state yields an excitation energy of around $153(14)$~keV for the $(8^-$) isomer. This preliminary value, which should be confirmed in future measurements, agrees with the suggested excitation scheme shown in Fig. 4.(a) of Ref.~\cite{Sto89}. There, the $(8^-)$ isomer lies just below the $(5^+)$ level at $162.8$~keV. Contrary to the general trend, the lower spin $(4^+)$ ground state was produced in greater quantities than the higher-spin isomer. In future, the measured states should be identified by measuring their beta-decay half-lives after the trap. 

\section{Summary and outlook}
\label{sec:summary}

In this work, we have utilized JYFLTRAP Penning trap for measuring the energies of the $(11/2^-)$ neutron-hole states in $^{121,123,125}$Cd and $(1/2^-)$ proton-hole states in $^{129,131}$In. In addition, the energies of the ($7^-$) isomers in $^{130}$Sn and $^{134}$Sb were measured and an agreement with the precisely known adopted values was found. The new excitation energy of the isomer in $^{123}$Cd suggests that the decay scheme of Ref.~\cite{Huc89b} should be revised. For $^{131}$Sn, resolving the ground and isomeric states was not possible in this work. In order to identify the measured state, laser or post-trap spectroscopy would be useful in the future. In addition, an isomeric state observed in $^{132}$Sb should be verified and measured more precisely. The long-lived high-spin isomers in $^{129}$In and $^{131}$In should also be searched for in forthcoming mass measurements.

The results of this work providing precise information on energy levels around doubly magic $^{132}$Sn are important for the description of these nuclei within the nuclear shell model. Moreover, the exact knowledge of the masses of the ground and isomeric states is crucial for the modeling of the astrophysical $r$-process which proceeds via the region of the studied nuclei. The new excitation energies determined in this work are more precise than the previously adopted values and large discrepancies have been found for some isomers. These new data should be taken into account in future calculations. 

\begin{acknowledgments}
This work has been supported by the Academy of Finland under the Finnish Centre of Excellence Programme 2006-2011 (Nuclear and Accelerator Based Physics Programme at JYFL). A.K. acknowledges the support from the Academy of Finland under the project 127301. The authors would like to thank Dr. Gabriel Mart\'inez-Pinedo for discussions concerning the role of isomeric states in the $r$-process.
\end{acknowledgments}

\end{document}